\documentclass{an}
\usepackage{graphicx}
\usepackage{times}

\def\Feii{[Fe{\textsc{ii}}]\,}
\def\etal{\emph{et\,al.\,}}
\def\HA{H{\small{$\alpha$}}\,}
\def\HB{H{\small{$\beta$}}\,}
\def\PaA{Pa{\small{$\alpha$}}\,}
\def\PaB{Pa{\small{$\beta$}}\,}

\Volume{00}              
\Year{0000}              
\Month{00}               
\Pagespan{000}{000}      

\begin{document}
\lhead[\thepage]{R.G. Sharp : Age-dating a star-burst with
GEMINI/CIRPASS observations of the core of M83}
\rhead[Astron. Nachr./AN~{\bf XXX} (200X) X]{\thepage}
\headnote{Astron. Nachr./AN {\bf 32X} (200X) X, XXX--XXX}

\title{Age-dating a star-burst with GEMINI/CIRPASS observations of the core of M83}

\author{R.G. Sharp\inst{1},
I.R. Parry\inst{1},
S.D. Ryder\inst{2},
J.H. Knapen\inst{3} and
L.M. Mazzuca\inst{4}
}
\institute{
Institute of Astronomy, University of Cambridge, Madingley Road, Cambridge, CB3 0HA, UK
\and
Anglo-Australian Observatory, P.O. Box 296, Epping , NSW 1710, Australia 
\and
University of Hertfordshire , College
Lane, Hatfield, Hertfordshire AL10 9AB, UK
\and
NASA Goddard Space Flight Center
}

\date{Received {date will be inserted by the editor}; 
accepted {date will be inserted by the editor}} 

\abstract{We present preliminary results from a set of near-IR
integral field spectroscopic observations of the central, star-burst,
regions of the barred spiral galaxy M83, obtained with CIRPASS on
Gemini-S. We present maps in the \PaB and \Feii\,1.257\,$\mu$m
emission lines which appear surprisingly different.  We outline the
procedure in which we will use \PaB emission line strengths and
measures of CO absorption to determine the relative and absolute ages
of individual star-forming knots in the central kpc region of M83.
 }

\correspondence{Robert Sharp rgs@ast.cam.ac.uk}

\maketitle

\section{Introduction}
The central star-burst in M83 (NGC 5236) has been studied
photometrically by Harris \etal (2001) using \emph{HST/WFPC} images in
the broad-band near-UV and optical, as well as narrow-band \HA and \HB
to derive colours and line equivalent widths for 45 clusters. Despite
the excellent spatial resolution of these observations, optical
photometric analyses such as these suffer from patchy dust extinction,
a reddening vector which parallels the evolutionary tracks in the
two-colour diagram, and selection effects which tend to exclude the
very youngest (\emph{t}$<$5\,Myr) clusters which have strong emission
lines, but only a weak stellar continuum.  Additionally, it is not
possible, on the basis of broad-band colours alone, to distinguish an
instantaneous burst of star formation from a constant star formation
rate.

To help overcome these effects, we use the Integral Field Unit (IFU)
of the Cambridge Infra-Red Panoramic Survey Spectrograph (CIRPASS;
Parry \etal 2000)\footnote{http://www.ast.cam.ac.uk/$\sim$~optics/} to
measure both the \PaB equivalent width, as well as the CO(6,3)
spectroscopic index over the entire star-forming arc and nucleus of
M83. As demonstrated by Ryder \etal (2001) for M100, the
combination of two such diagnostics allows one to constrain not just
the age, but also the burst duration, for each cluster.

\begin{figure}
\centering
\includegraphics[width=8cm]{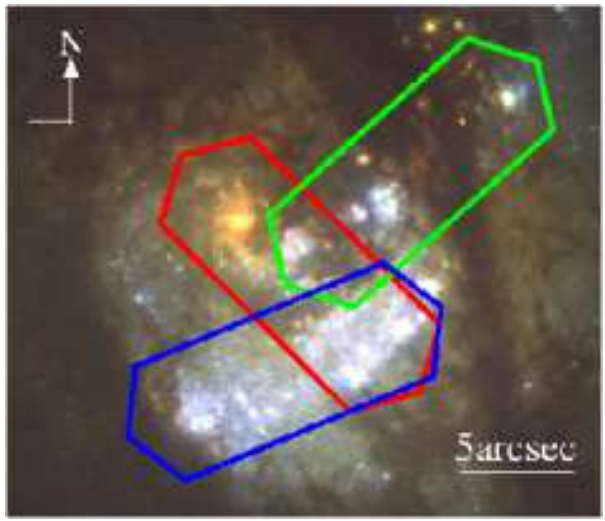}
\caption{\label{M83 IFU map} \emph{HST} colour composite image from
Harris \etal 2001 (F300W, F547M and F814W) overlaid with the outline
of the CIRPASS IFU indicating the three telescope pointings targeted.
The optical galaxy nucleus is the bright spheroidal knot to the
North-East of the centre of the image, at the end of the central IFU
pointing.  A naive interpretation of this three colour image suggests
that much of the star formation activity has taken place in the
southern arc of the nucleus, with isolated pockets of activity
elsewhere.  Using IFU spectroscopy observations one can survey the
entire nuclear region, looking for the signatures of star-formation,
past and present, without the bias inherent in previous long slit
studies which have, by necessity, only targeted regions highlighted by
broad band imaging techniques.}
\end{figure}

\section{The CIRPASS spectrograph}
CIRPASS is a fibre-fed IR \emph{J+H} band spectrograph with
Multi-Object (MOS) and Integral Field (IFU) modes.  The cryogenic
camera is liquid nitrogen cooled while the spectrograph optics are
housed in a refrigerated cold room cooled to -42$^{\circ}$\,C.  The
fibre-fed design introduces considerable flexibility of operation for
CIRPASS, allowing observations on a range of telescopes (from the 8
inch laboratory test telescope used for system verification, the 4m
class Anglo-Australian and William Hershel telescopes to the 8m Gemini
south telescope) with minimal modification to the focal plane
interface.

In MOS mode 150 independent fibres of $\sim$1.6\,arcsec diameter are
available over a field-of-view (FoV) of 17$\times$17\,arcmin to
40$\times$40\,arcmin (fibre aperture and FoV are dependent on the
details of the telescope focal station occupied).  Fibres are
typically deployed in pairs to target up to 75 objects at a time while
simultaneously sampling the sky.  The telescope is then nodded between
the two fibres of each pair to allow sky subtraction.

\begin{figure}
\centering
\includegraphics[width=7cm]{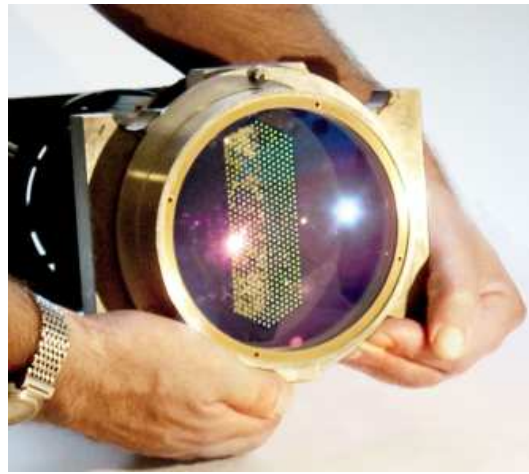}
\caption{\label{IFU}Here we see the Integral Field Unit (IFU) of
CIRPASS.  This unit is attached at the focal plane of the telescope.
The broadly rectangular array of fibres can be seen through the field
lens.}
\end{figure}

CIRPASS performs integral field spectroscopy using an array of 490
macrolenses which make up the Integral Field Unit (IFU), shown in
Figure \ref{IFU}.  A macrolens array is used in favour of a microlens
array to reduces losses from Focal Ratio Degradation (FRD, Parry \etal
2000).  Two broadly rectangular fields of view are available,
13$\times$4.7\,arcsec and 9.3$\times$3.5\,arcsec.  The respective lens
scales are 0.36\,arcsec and 0.25\,arcsec.  Finer scales are possible
if taking advantage of a telescope equipped with an adaptive optics
system.  In IFU mode CIRPASS is typically operated in medium
resolution mode (\emph{R}$\sim$5000) giving a dispersion of
$\sim$2.2\,\AA/pixel.  Such a resolution allows CIRPASS to capitalise
on the intrinsically dark sky background between the OH air-glow lines
(Martini \& DePoy 2000), which are responsible for much of the high
sky brightness observed in the near-IR

Prior to September 2003, an Hawaii 1K HgCdTe array has been used in
the CIRPASS camera, giving a single shot spectral range of
$\sim$2200\,\AA\ at a two pixel resolution of 4.4\AA.
The detector was upgraded to an Hawaii 2K array in September 2003 and
has been used successfully for science during CIRPASS MOS mode
observations at the Anglo-Australian Telescope (AAT) during October
2003.  The larger detector provides complete coverage of the \emph{J}
or \emph{H} bands, between the atmospheric water absorption troughs,
in single observations. Alternatively, many programms may benefit from
increased resolution without the loss of wavelength coverage that
would be inherent with the 1K array.

\section{Observations of the nuclear region of M83}
During March 8$^{\rm{th}}$ and 9$^{\rm{th}}$ 2003, observations were
undertaken of M83 (NGC 5236), with CIRPASS in IFU mode at the
cassegrain focus of the 8m Gemini-South telescope.  Both nights were
photometric with sub-arcsecond seeing.  During \emph{H} band
observations, marginally increased variability was observed in OH sky
emission features, resulting in sub-optimal sky subtraction.  This is,
however, not believed to compromise the observational dataset.  Three
overlapping regions, close to the optical nucleus of M83, were
observed (see Figure \ref{M83 IFU map}).  The 0.36\,arcsec lens scale
was used giving a FoV per pointing of $\sim$13$\times$4.7\,arcsec.
Observations where obtained in both the \emph{J} and \emph{H} bands
centred at $\lambda_{\rm{cen}}=12530$\,\AA\ and
$\lambda_{\rm{cen}}=15590$\,\AA.  The wavelength ranges cover the \PaB
and \Feii\,1.257\,$\mu$m emission lines in \emph{J} and the
\Feii\,1.644\,$\mu$m line and CO(6,3) molecular absorption feature in
the \emph{H} band.  Two 900\,sec background limited exposures were
taken in each band at each location, interleaved with 900\,sec
off-target observations for sky subtraction.  Observations are
performed using a Non-Destructive Read (NDR) mode.  Three sequences of
10 detector reads were done at 5\,min intervals throughout the
integration, to facilitate cosmic ray rejection and to suppress
read-noise.  A small dither of $\sim$0.36\,arcsec (one IFU lens) was
performed between on-target frames to allow the removal of the small
number of detector defects.

\subsection{Data Reduction}
The data have been processed using the \textsc{cirpass iraf}
package\footnote{The \textsc{cirpass iraf} package can be obtained
from http://www.ast.cam.ac.uk/$\sim$optics/cirpass/docs.html}.  A data
reduction \emph{cookbook} detailing the package tasks is under
development\footnote{A data reduction guide for the \textsc{cirpass
iraf} can be found at
http://www.ast.cam.ac.uk/$\sim$optics/cirpass/datared/cookbook\_sn1987a.php}.
Data reduction proceeds in the usual manner.  A reset frame
correction, analogous to bias subtraction for a CCD frame, is
performed.  Cosmic ray rejection is carried out for the 900\,sec
frames using the \textsc{iraf imcombine} task and \textsc{ccdclip}
pixel rejection to combine the time series data obtained during each
NDR observation.  A pixel-to-pixel flat field correction and a bad
pixel mask are derived from non-dispersed frames with the detector
illuminated by a low power tungsten lamp.  Interleaved sky frames are
then subtracted from target frames, this step removing the need for
the subtraction of a \emph{dark frame}.  The spectra are extracted
using the optimal extraction routine described by Johnson \etal
(2003).  Spectra are also extracted from a tungsten illuminated dome
flat field frame.  The 2D spectral images are then divided by the
extracted 2D flat field frame to account for fibre-to-fibre variation.
A wavelength solution is derived from the the co-added sky frames
using the OH air-glow lines using the line list of Maihara \etal
(1993).  The spectra are then transformed to a common wavelength scale
using the \textsc{iraf longslit} package tasks \textsc{fitcoords} and
\textsc{transform} and pairs of dither positions are mosaiced.  Flux
calibration and removal of telluric absorption features is performed
using the spectrum of an F2V star, HD128299, observed at an airmass
similar to the target observations.  The stellar spectrum is fitted by
the appropriate black body spectrum (\emph{T}$_{\rm{eff}}$=6750K).
Data analysis is performed within an interactive software suite
written in the IDL programming environment.

\subsection{Image registration}
The \emph{J} and \emph{H} band data were recorded on different nights.
In order to register the data a continuum image is created,
reconstructing the IFU image across the full wavelength range in each
band.  Primary features in the final continuum image are the southern
star-forming arc and the optical nucleus.  Using these features we can
register the data to within the individual IFU lens resolution
($\sim$0.36\,arcsec).

\subsection{Variance array propagation}
The two dimensional spectra are critically sampled on the detector
with a FWHM of $\sim$1.8\,pixels.  The peak-to-peak separation of
adjacent spectra is $\overline{y}=2.0$\,pixels.  In order to extract
spectra with such close separations, leading to a significant overlap
between the wings of adjacent spectral profiles, an optimal extraction
technique is implemented solving for the maximum likelihood light
distribution of each fibre triplet (a fibre and its two nearest
neighbours).  The central position of each fibre, relative to a set of
observations of well spaced ($\overline{y}>100$\,pixels) calibration
fibres observed at each instrument setting prior to the science
exposures, is accurately known from extensive laboratory tests.

This process (and that of accurate emission line profile fitting)
requires the propagation of the expected variance associated with each
pixel of the detector.  With 900\,sec exposures, the noise in CIRPASS
observations is dominated by the count rate from thermal background
emission below 1.67\,$\mu$m.  A short pass blocking filter prevents
thermal emission long-ward of 1.67\,$\mu$m entering the camera optics.
At the small number of wavelengths affected by strong OH air-glow
lines, noise is significantly enhanced by Poisson noise from the high
intensity lines.

\subsection{Equivalent width measurements}
We fit profiles to the three prominent emission lines: \PaB,
\Feii\,1.257\,$\mu$m and \Feii\,1.644\,$\mu$m using a reduced $\chi^2$
technique and assuming a single Gaussian profile.  Such a simple
profile is believed to be appropriate at the level of spatial and
spectral resolution.  During line fitting a spectral mask is used to
weight the fit against pixels affected by strong OH air-glow line
residuals.  This OH avoidance technique is a fundamental source of the
sensitivity of CIRPASS.

Origlia, Moorwood \& Oliva (1993) describe the CO(6,3) absorption band
and provide a prescription for measuring the equivalent width of the
feature.  They define the local continuum using a linear fit to two
continuum points which straddle the CO band.  The equivalent width is
then integrated over a predefined pass band.  The width of this band
(1.6175\,$\mu$m-1.6220\,$\mu$m, 45\,\AA\ or 20\,pixels) is of the
order of the velocity structure seen in the nuclear regions of M83
($\pm$150\,km\,s$^{-1}$, $\pm$8\,\AA, $\pm$3.6\,pixels).  It is
therefore not desirable to simply define a single systemic
velocity-adjusted pass band for the entire observational data set.  To
overcome this we define a smooth background velocity field based on
the observed velocity of the \PaB emission line.

\begin{figure*}
\centering
\includegraphics[width=8cm]{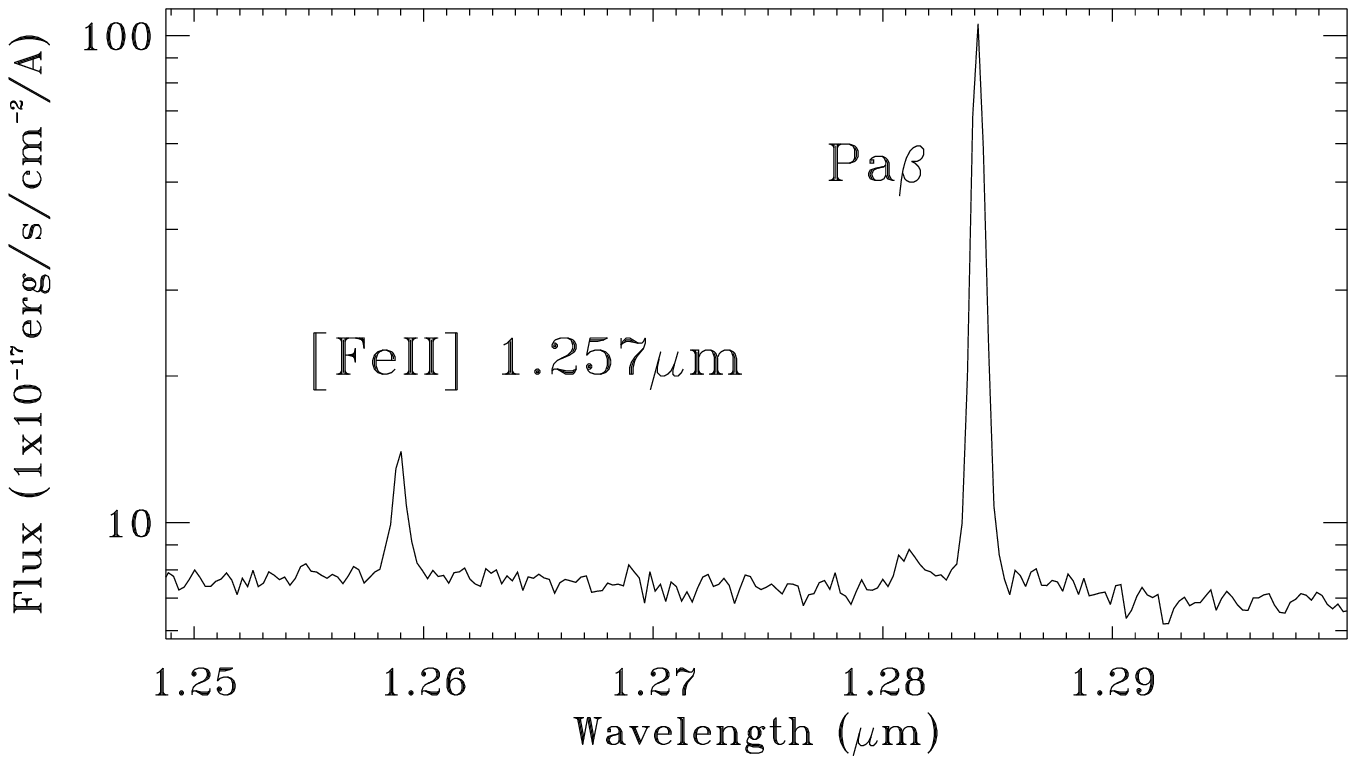}
\vspace{1cm}
\includegraphics[width=8cm]{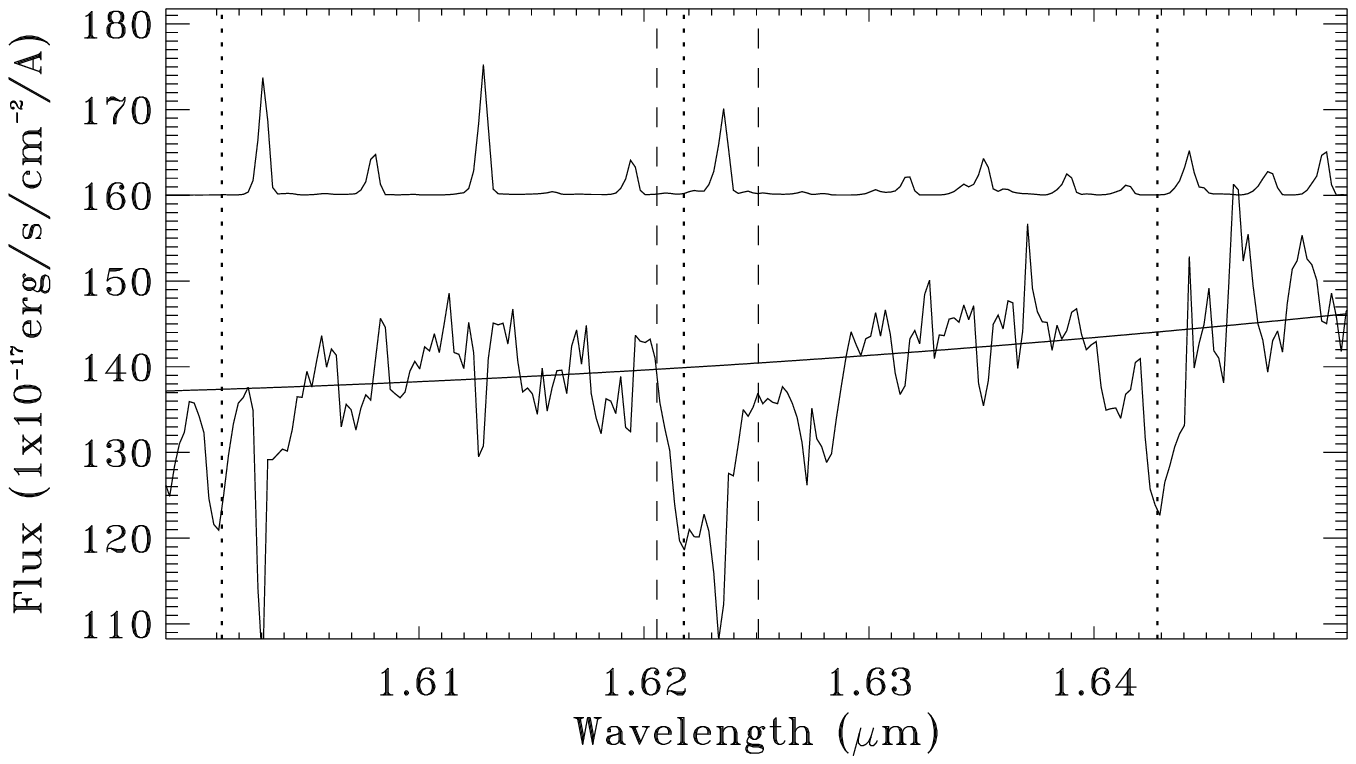}
\caption{\label{spectra}Two example spectra are shown.
\newline
Left : In the \emph{J} band we see the strong \PaB emission, a
signature of hot young stars and active star formation, along with the
\Feii\,1.257\,$\mu$m line.  This spectrum is taken from a knot of
strong hydrogen emission.  While the knot is visible as diffuse
emission in \emph{HST/}NICMOS \PaA imaging, it is not evident in broad
band continuum observations at other wavelengths.
\newline
Right : The CO(6,3) molecular absorption features are clearly visible
in this \emph{H} band spectrum taken from the bright optical nucleus
of the galaxy (lower spectrum).  The location of the centres of three
of the CO band heads, after correction for redshift, are marked by
dotted vertical lines.  Two dashed vertical lines indicate the extent
of the CO(6,3) band as defined by Origlia, Moorwood and Oliva (1993).
A polynomial fit to the continuum level is also marked.  The upper
spectrum shows the OH air-glow emission spectrum from which a mask is
created to weight spectral fitting algorithms.  The CO features are
believed to arise in the atmospheres of cool G/K and M red giant stars
and can be used, in combination with the \PaB emission, to constrain
the age and burst duration of star formation in this region.}
\end{figure*}

\section{Summary of results}
\label{summary}
Quantitative analysis is still in progress, but already we see
evidence for a much more complex star formation history than that
hinted at by the single long-slit near-IR spectrum of Thatte \etal
(2000).

Figure \ref{spectra} shows two example spectra. The left hand spectrum
is taken from a region visible in narrow band \emph{HST}/NICMOS \PaA
imaging with extended emission but which does not show signs of active
star formation loci in broad band WFPC2/NICMOS imaging.  This
highlights one of the key goals of these observations, to perform an
accurate census of the star formation in the nuclear region of M83,
without bias towards bright continuum regions, such as those targeted
by Harris \etal (2001).  In contrast to the strong \PaB emission,
little CO absorption is seen in the \emph{H} band spectrum of this
region (not shown in Figure \ref{spectra}), suggesting a young
($<5$\,Myr) star formation region.

The right hand spectrum of Figure \ref{spectra} is taken from the
optical nucleus of M83.  Strong CO absorption is visible in this
\emph{H} band spectrum which, when taken in comparison with a very
weak \PaB emission line strength in the \emph{J} band spectrum (not
shown), is suggestive of an old stellar population ($>15$\,Myr).  By
comparing the equivalent width ratio for \PaB and CO(6,3) we will
determine the age of star-formation across the nucleus of M83.

\begin{figure*}
\centering
\includegraphics[width=5cm]{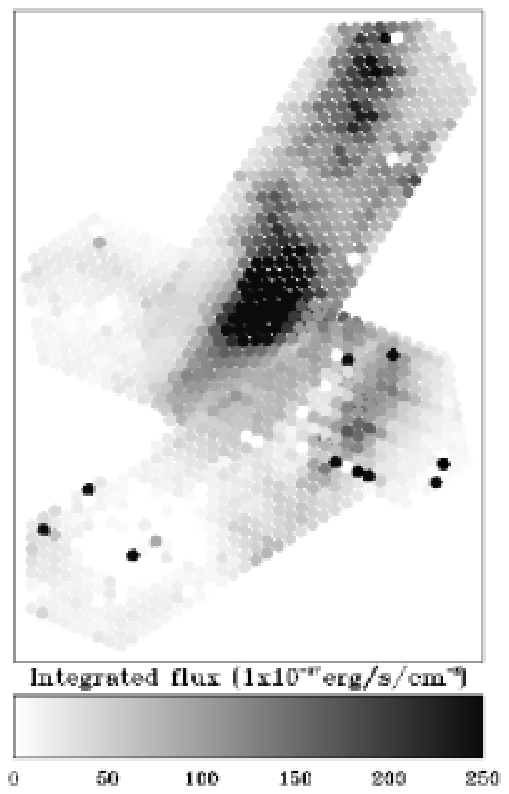}
\hspace{1cm}
\includegraphics[width=5cm]{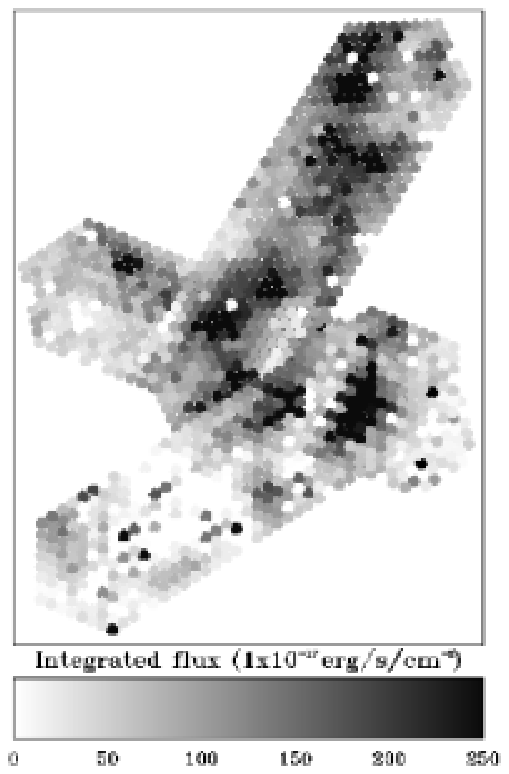}
\caption{\label{PaB FeII}A comparison of the \PaB line flux (left),
with that of the \Feii\,1.257\,$\mu$m line (right) is made above. See
discussion in Section \ref{summary}.}
\end{figure*}

Figure \ref{PaB FeII} compares the distribution of \PaB line flux,
with that of the \Feii\,1.644\,$\mu$m.  The former provides a measure
of the massive star formation rate within the last $\sim$5\,Myr, while
the latter is thought to be a tracer of shocks from supernova
explosions within the past 10$^4$\,years (Alonso-Hererro \etal
2003). The \Feii emission is rather clumpy, and confined mainly to the
periphery of the more diffuse \PaB emission, suggesting that massive
star formation is being propagated into regions of undisturbed gas by
the passage of the supernova blasts themselves.

\acknowledgements

The authors wish to thank and acknowledge the following for their
contributions to this work. Firstly, we thank the Instrumentation
Group at the Institute of Astronomy, Cambridge for providing
CIRPASS. We also thank the CIRPASS team for support and operation of
the instrument at the telescope. The Raymond and Beverly Sackler
Foundation and PPARC were responsible for the funding of
CIRPASS. Finally we acknowledge the Gemini Observatory for the
allocation of telescope time, allowing CIRPASS to be used as a visitor
instrument, supporting its installation at the telescope and operating
the telescope while the observations were made.

The Gemini Observatory is operated by the Association of Universities
for Research in Astronomy, Inc., under a cooperative agreement with
the NSF on behalf of the Gemini partnership: the National Science
Foundation (United States), the Particle Physics and Astronomy
Research Council (United Kingdom), the National Research Council
(Canada), CONICYT (Chile), the Australian Research Council
(Australia), CNPq (Brazil) and CONICET (Argentina)


\begin{thebibliography}{}
\bibitem{} Alonso-Hererro A. \etal : 2003, AJ, 125, 1210
\bibitem{} Harris J. \etal : 2001, ApJ, 122, 3046
\bibitem{} Johnson R. \etal : 2003, in prep.
\bibitem{} Martini P. and DePoy D.L. : 2000, SPIE, 4008, 695
\bibitem{} Maihara T. \etal : 1993, PASP, 105, 940
\bibitem{} Origlia, Moorwood A.F.M., and Oliva E. : 1993, A\&A 280 536
\bibitem{} Parry I. \etal : 2000, SPIE, 4008, 1193 
\bibitem{} Ryder S.D., Knapen J.H. and Takamiya M. : 2001, MNRAS, 323, 663
\bibitem{} Thatte N., Tecza M. and Genzel R. :  2000, A\&A, 364, L47
\end{thebibliography}
\end{document}